\documentclass[12pt]{iopart}
\pdfoutput=1

\usepackage{graphicx}
\usepackage{iopams}  
\usepackage{url}

\begin{document}

\title{A Quantitative Approach to Painting Styles}

\author{Vilson Vieira$^1$, Renato Fabbri$^1$, David Sbrissa$^1$, Luciano da
  Fontoura Costa$^1$, Gonzalo Travieso$^1$}

\address{$^1$ Instituto de F\'isica de S\~ao Carlos, Universidade de S\~ao Paulo
  (IFSC/USP)}
\eads{\mailto{vilsonvieira@usp.br}, \mailto{fabbri@usp.br},
  \mailto{davidsbrissa@hotmail.com}, \mailto{ldfcosta@gmail.com}, \mailto{gonzalo@ifsc.usp.br}}

\begin{abstract}
  This research extends a method previously applied to music and philosophy~\cite{vieira},
  representing the evolution of art as a time-series where relations like
   \textit{dialectics} are measured
  quantitatively. For that, a corpus of paintings of 12 well-known artists from
  baroque and modern art is analyzed. A set of 93 features is extracted and
  the features which most contributed to the classification of painters are
  selected. The projection space obtained provides the basis
  to the analysis of measurements. This quantitative measures underlie revealing
  observations about the evolution of painting styles, specially when compared with other
  humanity fields already analyzed: while music evolved along a
  master-apprentice tradition (high dialectics) and philosophy by opposition,
  painting presents another pattern: constant increasing skewness, low
  opposition between members of the same movement and opposition peaks in the
  transition between movements. Differences between baroque and modern
  movements are also observed in the projected ``painting space'': while baroque
  paintings are presented as an overlapped cluster, the modern paintings
  present minor overlapping and are disposed more widely in the projection
  than the baroque counterparts. This finding suggests that baroque painters shared aesthetics while modern painters tend to ``break rules'' and develop their own style.
\end{abstract}

\vspace{2pc}


\maketitle

\section{\label{sec:introduction}Introduction}

Painting classification is a common field of interest for applications such as
painter identification --- e.g.\ assessing the authenticity of a given art work
--- style classification, paintings data base search and more recently,
automatic aesthetic judgment in computational creativity
applications. Determining the best features for painting style characterization
is a complex task on its own. Many studies~\cite{ana,icoglu,spehr,johnson} applied
image processing to feature extraction for painter and art movements
identification. Manovich~\cite{manovich,manovich2,manovich3} uses features like
entropy, brightness and saturation to map paintings and general images into a
2-dimensional space and, in this way, to visualize the difference between
painters. There are also many related
works dealing on feature selection for painting classification. Penousal et
al.~\cite{penousal} use features based on aesthetic criteria estimated by image
complexity while Zujovic et al.~\cite{zujovic} evaluate a large set of features
that most contribute to classification.

This study also analyses a set of features which most contribute to the classification of paintings. Although, in contrast with previous works, it goes forward: the historic evolution of painting styles is analyzed by means of geometric measures in the feature space. Those measures -- \textit{opposition}, \textit{skewness} and \textit{dialectics} -- are central while discussing human history. However, such discussions are common only at humanities fields like Philosophy and those quantitative measures are suggested to do not surpass but contribute in this understanding of human history.

To create the feature space, a set of 93 features is extracted from 240 images of 12 well-known
painters. The first six painters of this group represent the baroque movement
while the remaining six represent the modern art period. A feature selection
process yields the pair of features which most contributed for the
classification. Similar results using LDA (Linear Discriminant Analysis)
analysis are obtained, which reinforces the feature selection.

After feature selection, a centroid for each group of paintings is calculated which defines a \textit{prototype}: a representative work-piece for the respective cluster. The set of all prototypes following a chronological order defines a time-series where the main purpose of this study is performed: the
quantitative analysis of the historical evolution of art movements. Extending a
method already applied to music and philosophy~\cite{vieira}, \textit{opposition},
\textit{skewness} and \textit{dialectics} measurements are taken. These concepts are central in philosophy --- e.g.\ philosophers from antiquity like Aristotle and Plato developed their ideas using the dialectics method while it is also found in modern works like Hegelian and Marxist dialectics --- and humanistic fields, however lacks studies from a quantitative perspective.~\cite{dialectics} Represented as geometric measures, these concepts reveal interesting results
and patterns. Modern paintings groups show minor superposition when compared with baroque counterparts suggesting the independence in style found historically in modernists and strong influence of shared painting techniques found in baroque painters. Dialectics and opposition values presented a peak in the transition between baroque and modern periods --- as expected considering history of art --- with decreasing values in the beginning of each period. Skewness index is presented with oscillating but increasing values during all the time-series, suggesting a constant innovation through art movements. These results present an interesting counterpart with previous results in philosophy --- where opposition is strong in almost entire time-series --- and in music --- where the dialectics is remarkable~\cite{vieira}.

The study starts describing the corpus of paintings used and a review of both aesthetic and historic facts regarding baroque and modern movements (Section~\ref{sec:methodology}). The image processing steps used to extract features from these paintings are presented followed by the feature selection. The results are them discussed in Section~\ref{sec:results} with basis on geometric measurements in the projected feature space -- considering the most clustered projection and LDA components. 
 
\section{\label{sec:methodology}Modeling painting movements}

\subsection{Painting corpus}

A group of 12 well-known painters is selected to represent artistic styles or
movements from baroque to modernism. Six painters are chosen to represent each
of these movements.  The group is presented in Table~\ref{tab:painters} together
with their more representative style, in chronological order.  It is known that
painters like Picasso covered more than one style during his life. Although,
only the most remarkable style is selected intending to well characterize the
painter by means of this specific period or movement.

\begin{table}
\caption{\label{tab:painters} Painters ordered
  chronologically with the artistic style they represents.}
\begin{indented}
\item[] \begin{tabular}{@{}ll}
\br

 artists                       & Remarkable Styles/Movements \\ 
 
 \mr

 Caravaggio                    & Baroque, Renaissance \\
 Frans Hals                    & Baroque, Dutch Golden Age \\
 Nicolas Poussin               & Baroque, Classicism \\
 Diego Vel\'{a}zquez           & Baroque \\
 Rembrandt                     & Baroque, Dutch Golden Age, Realism \\
 Johannes Vermeer              & Baroque, Dutch Golden Age \\
 
 \mr
 
 Vincent van Gogh              & Post-Impressionism \\
 Wassily Kandinsky             & Expressionism, Abstract art \\
 Henri Matisse                 & Modernism, Impressionism \\
 Pablo Picasso                 & Cubism \\
 Joan Mir\'{o}                 & Surrealism, Dada \\
 Jackson Pollock               & Abstract expressionism \\

\br
\end{tabular}
\end{indented}
\end{table}

For each painter, 20 raw images are considered from the database of public
images organized by Wikipedia. Examples of selected paintings titles and their
respective creation year are listed in Table~\ref{tab:paintings}\footnote{The
  source code together with all the 240 raw images are available online at
  \url{http://github.com/automata/ana-pintores.}}

\begin{table} 
  \caption{\label{tab:paintings} Some of the 240 selected paintings and their
    respective author and year of creation.}
\begin{indented}
\item[] \begin{tabular}{@{}lll}

\br

 Painter & Painting title                       & Year \\ 
 
 \mr

 Caravaggio          & Musicians & 1595 \\
                     & Judith Beheading Holofernes & 1598 \\
                     & David with the Head of Goliath & 1610 \\
                      
 Frans Hals          & Portrait of an unknown woman & 1618/20 \\
                     & Portrait of Paulus van Beresteyn & 1620s \\
                     & Portrait of Stephanus Geeraerdts & 1648/50 \\
 Nicolas Poussin     & Venus and Adonis & 1624 \\
                     & Cephalus and Aurora & 1627 \\
                     & Acis and Galatea & 1629 \\
 
 Diego Vel\'{a}zquez & Three musicians & 1617/18 \\
                     & The Lunch & 1618 \\
                     & La mulatto & 1620 \\
 
 Rembrandt           & The Spectacles-pedlar (Sight) & 1624/25 \\
                     & The Three Singers (Hearing) & 1624/25 \\
                     & Balaam and the Ass & 1626 \\
Johannes Vermeer    & The Milkmaid & 1658 \\
                    & The Astronomer & 1668 \\
                    & Girl with a Pearl Earring & 1665 \\
                    
\mr
                    
Vincent van Gogh    & Starry Night Over the Rhone & 1888 \\
                    & The Starry Night & 1889 \\
                    & Self-Portrait with Straw Hat & 1887/88 \\
Wassily Kandinsky   & On White II & 1923 \\
                    & Composition X & 1939 \\
                    & Points & 1920 \\
Henri Matisse       & Self-Portrait in a Striped T-shirt & 1906 \\
                    & Portrait of Madame Matisse & 1905 \\
                    & The Dance (first version) & 1909 \\
Pablo Picasso       & Les Demoiselles d'Avignon & 1907 \\
                    & Guernica & 1937 \\
                    & Dora Maar au Chat & 1941 \\
Joan Mir\'{o}       & The Farm & 1921/22 \\
                    & The Tilled Field & 1923/24 \\
                    & Bleu II & 1961 \\
Jackson Pollock     & No. 5 & 1948 \\
                    & Autumn Rhythm & 1950 \\
                    & Blue Poles & 1952 \\

\br
\end{tabular}
\end{indented}
\end{table}

It is interesting to raise some historical and aesthetic characteristics from
baroque and modern movements before entering the quantitative analysis in
Section~\ref{sec:results} where those hypothesis are further
discussed. Baroque is marked by tradition, a desire to portrait the truth (found
in Caravaggio, Frans Hals and Vel\'{a}zquez), the beauty (Poussin, Vermeer), the
nature and the sacred (Caravaggio, Rembrandt). A remarkable use of light
contrast (as in the ``\textit{chiaroscuro}'' technique mastered by Caravaggio),
disregarding for simple equilibrium in composition and preference for complex
oppositions, both compound aesthetic characteristics which baroque artists
used to represent their view of nature. The transmission of those techniques from
one painter to another is common in baroque.
Modernists, on the other hand, did not follow
``rules''. Each modern painter employed or created new ways to represent
nature. As noted by Gombrich~\cite{gombrich}: ``[they] craved for an art that
does not consists of tricks that could be learn, for a style that is not a mere
style, but something strong and powerful like the human passion''. Van Gogh
pursued this artistic trend in his intense use of colors and the caricature aspect of his paintings.
Paul Gauguin searched for ``primitive'' in his
paintings. Others, like Seurat, applied physical properties of the chromatic
vision and started painting the nature like a collection of color points, and
ended creating the pointillism. Modernists created a new style for each of their
experiments using their own techniques to represent a nature outside of
the domains already covered by their predecessors.

\subsection{Image processing}

\begin{figure}[ht!]
    
\begin{center}
{\centering
        \includegraphics[scale=.6]{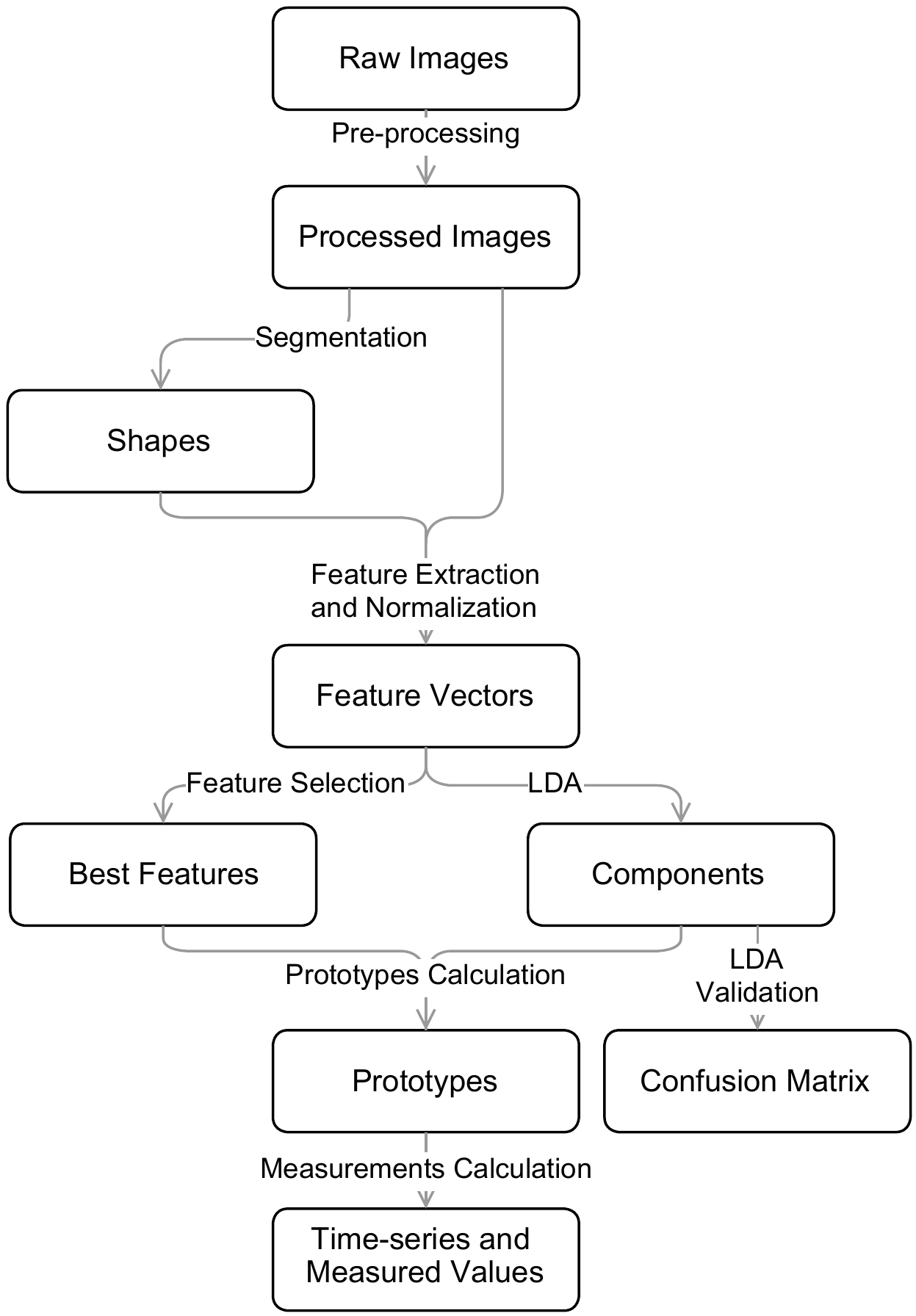}}
      \caption{A summary of all steps from image processing through
        feature extraction through time series and measurements calculation (skewness,
        opposition and dialectics).}
        \label{fig:dataflow}
  \end{center}
\end{figure}

All 240 images are re-sized to 800x800 pixels and cropped to consider a region
positioned in the same coordinates and with same aspect of both original
paintings, and pre-processed by applying histogram equalization and median
filtering with a 3-size window. Feature extraction algorithms are applied to
colored, gray-scale or binary versions of images as necessary (e.g.\ convex-hull
used a binary image, whereas Haralick texture used the gray-scale image and SLIC
segmentation analysis is applied to color images). Curvature measurements are
extracted from segments of paintings identified by the SLIC segmentation
method~\cite{slic} as presented in Figure~\ref{fig:curvatura}. The whole process is represented schematically in
Figure~\ref{fig:dataflow} and covers all the steps from image processing through
measurements, discussed in the following sections.

\subsection{Extracted features}

To create a \textit{painting space} a number of
distinct features extracted by computational methods from raw images of the
paintings is considered. The features are related with aesthetics
characteristics and aim to quantify properties well-known by art critics. All
the features are summarized in Table~\ref{tab:features} and detailed, grouped in classes, in the
following list.

\begin{table}[ht]
\caption{\label{tab:features} Extracted features.}
\begin{indented}
\item[] \begin{tabular}{@{}ll}
\br

 Number of features            & Features \\ 
 
 \mr

 4                             & Energy of the whole image \\
 4                             & Energy $\mu$ of image rows \\
 4                             & Energy $\sigma$ of image rows \\
 4                             & Energy $\mu$ of image columns \\
 4                             & Energy $\sigma$ of image rows \\
 4                             & Energy centroids of image rows \\
 4                             & Energy centroids of image columns \\
 4                             & Energy $\mu$ of rows and columns \\
 4                             & Energy $\sigma$ of rows and columns \\
 
 \mr
 
 1                             & $\mu$ of local entropy (5-size window) \\
 1                             & $\mu$ of local entropy (50-size window) \\
 
 \mr

 4                             & Angular second moment \\
 4                             & Contrast \\
 4                             & Correlation \\
 4                             & Sum of squares: variance \\
 4                             & Inverse difference moment \\
 4                             & Sum average \\
 4                             & Sum variance \\
 4                             & Sum entropy \\
 4                             & Entropy \\
 4                             & Difference average \\
 4                             & Difference entropy \\
 
 \mr

 2                             & $\mu$ of distance between curvature peaks \\
 2                             & $\sigma$ of distance between curvature peaks \\
 1                             & $\mu$ of number of curvature peaks \\
 1                             & $\mu$ of segments perimeter \\
 1                             & $\mu$ of segments area \\
 1                             & $\mu$ of circularity ($Per.^2/Area$) \\
 1                             & $\mu$ of number of segments \\
 1                             & $\mu$ of convex-hull area \\
 1                             & $\mu$ of convex-hull and original areas ratio\\
 
 \mr
 
 93                            & Total of extracted features \\
 
 \br
 
\end{tabular}
\end{indented}
\end{table}

\textbf{\emph{General shape features}}: after image segmentation, a number of
shape descriptors are calculated for each segment, represented as a binary
matrix. \emph{Perimeter} is measured as pixel-length of the segment
contour. \emph{Area} is estimated counting the number of pixels representing the
segment. A convex-hull of the segment is used to calculate the \emph{convex
  area} and its ratio to the original segment area. The \emph{number of
  constituent segments} for each painting is also considered as a descriptor.

\textbf{\emph{Simple complexity features}}: \emph{Circularity} reveals how
much a shape remembers a circle and is obtained by the ratio between
perimeter and area of the segment. To estimate image complexity, a
number of \emph{entropy} measures of its energy (squared FFT coefficients) are
computed --- listed in the first quarter of Table~\ref{tab:features}. Together
with entropy, a more specific family of measurements is considered for texture
characterization: the 11 \emph{Haralick texture features}~\cite{haralick} are
calculated for this purpose.

\begin{figure}[h!]
\begin{center}
  \includegraphics[width=\columnwidth]{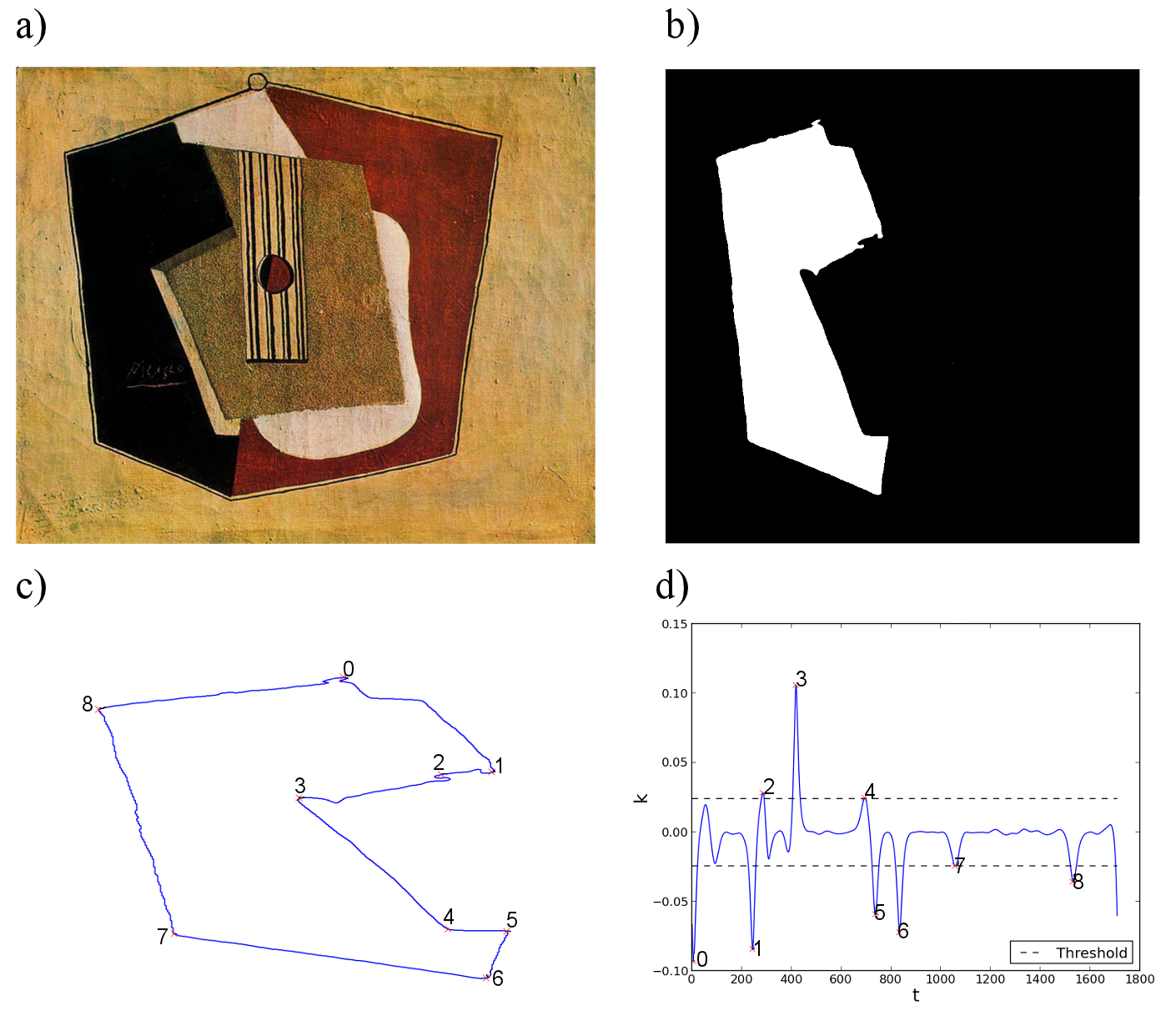}
      \caption{\textit{a)} The original paintings image. \textit{b)} A segmented
        region. \textit{c)} The extracted curvature of segment. \textit{d)} The
        parametric curve $k(t)$ with peaks given by a particular threshold.}
        \label{fig:curvatura}
        \end{center}
\end{figure}

\textbf{\emph{Curvature}}: this descriptor has an
interesting biological motivation related to the human visual system ---
e.g.\ object recognition is related to the identification of corners and high
curvature points~\cite{luciano}. Those points have more information about object shape than
straight lines or smooth curves. In this sense, curvature is well suited for the
characterization of the considered paintings. Curvature $k(t)$ of a parametric
curve $c(t) = (x(t), y(t))$ is defined as:

\begin{equation}
k(t) = \frac{\dot{x}(t)\ddot{y}(t)-\dot{y}(t)\ddot{x}(t)} {(\dot{x}(t)^2+\dot{y}(t)^2)^\frac{3}{2}}
\end{equation}

\noindent being $t$ the arc-length parameter and $\dot{x}(t)$, $\dot{y}(t)$, $\ddot{x}(t)$ and $\ddot{y}(t)$ are
respectively the first and second order derivatives of $x(t)$ and $y(t)$. Those
derivatives are obtained through Fourier transform and convolution theorem:

\begin{equation}
\dot{x} = \Im^{-1}(2\pi i \omega X(\omega))
\end{equation}
\begin{equation}
\dot{y} = \Im^{-1}(2\pi i \omega Y(\omega))
\end{equation}
\begin{equation}
\ddot{x} = \Im^{-1}(-(2\pi\omega)^2 X(\omega))
\end{equation}
\begin{equation}
\ddot{y} = \Im^{-1}(-(2\pi\omega)^2 Y(\omega))
\end{equation}

\noindent where $\Im^{-1}$ is the inverse Fourier transform, $X$ and $Y$ are the
Fourier transform of $x$ and $y$ respectively, $\omega$ is the angular frequency and $i$ is the imaginary
unit (see Figure~\ref{fig:curvatura}).

The corresponding features are calculated from the curvature data: the
\emph{mean} and \emph{standard deviation} of data, the \emph{number of peaks}
and the \emph{distance} (geometric and in pixels) between peaks. It is important
to note that a \emph{peak} is defined as a high curvature point. A point $a$ is
considered a peak if its curvature $k(a)$ satisfies the following criteria:

\begin{eqnarray}
	k(a) & > & k(a-1) \\
	k(a) & > & k(a+1) \\
    k(a) & > & \tau
\end{eqnarray}

\noindent being $\tau$ the corresponding threshold defined as

\begin{equation}
	median\left(k\right) \gamma 
\end{equation}

\noindent where $\gamma$ is a factor obtained empirically as values which reveal the desired level of curvature detail.

\subsection{Measurements}

$N$ features define a $N$-dimensional space, also called \textit{painting space}
 where the following measurements are
calculated~\cite{vieira}. For simplification, a prototype $\vec{p_i}$ is defined
for each class $C_p$. Each prototype summarizes a painting class, being its
\textit{centroid}: $\vec{p_i} = \frac{1}{N_p} \sum_{i=1}^{N_p} \vec{f_i}$
calculated in the projected space as well.

A sequence $S$ of $\vec{p_i}$ states defines a time series. The average state at
time $i$ of states $\vec{p_1}$ through $\vec{p_j}$ is defined as:

\begin{equation}
\vec{a_i} = \frac{1}{i}\sum_{j=1}^k\vec{p}_j
\end{equation}

The opposite state defines an opposition measure from $\vec{p_i}$ as

\begin{equation}
\vec{r}_i = \vec{p}_i + 2(\vec{a}_i - \vec{p}_i)
\end{equation}

\noindent and in this way an opposition vector can be defined:

\begin{equation}
\vec{D}_i=\vec{r}_i - \vec{p}_i.
\end{equation}

Knowing that any displacement from one state $\vec{p_i}$ to another
state $\vec{p_j}$ is defined as

\begin{equation}
\vec{M}_{i,j} = \vec{p}_j - \vec{p}_i
\end{equation}

\noindent it is possible to define an \emph{opposition index} to
quantify how much a prototype $p_j$ opposes $p_i$ (a displacement in
direction of $\vec{r_i}$) or emphasis $p_i$ (a displacement in
$-\vec{r_i}$ direction):

\begin{equation}
W_{i,j} = \frac{\left< \vec{M}_{i,j}, \vec{D}_i\right>}{||\vec{D}_i||^2}
\end{equation}

However, the movements in such \textit{painting space} are not restricted to
confirmation or refutation of ``ideas''. Alternative ideas can exist out of this
dualistic displacement. This is modeled as a \emph{skewness index} which
quantifies how much a prototype $p_j$ is innovative when compared with $p_i$:

\begin{equation}
s_{i,j} = \sqrt{\frac{|\vec{p}_i-\vec{p}_j|^2
          |\vec{a}_i-\vec{p}_i|^2 - 
          [(\vec{p}_i-\vec{p}_j) 
            (\vec{a}_i-\vec{p}_i)]^2}
        {|\vec{a}_i-\vec{p}_i|^2}}
\end{equation}

Another measure arises when considering three consecutive states at times $i,j$
and $k$. Being $p_i$ the thesis, $p_j$ the antithesis and $p_k$ the synthesis, a
\emph{counter-dialectics index} can be defined being

\begin{equation}
d_{i \rightarrow k} = 
      \frac{|\left< \vec{v}_j-\vec{v}_i,\vec{v}_k \right> + 
        \frac{1}{2}\left<\vec{v}_i-\vec{v}_j, \vec{v}_i+\vec{v}_j\right>|}
           {|\vec{v}_j-\vec{v}_i|}
\end{equation}

\noindent or, the distance between $p_k$ and the middle-line (or
middle-hyperplane for $N$-dimensional spaces) between $p_i$ and $p_j$. In other
words, a $p_k$ state with higher $d_{i \rightarrow k}$ is far from the synthesis
(low dialectics) and vice-versa.

\subsection{Feature selection}

To select the most relevant features a dispersion measure of the clusters is applied using
scatter matrices~\cite{luciano}. For all the $N$ paintings, considering all
possible combinations of feature pairs $F_{N, a}$ and $F_{N, b}$, the
$S_{b}$ (between class) and $S_{w}$ (within class) scatter matrices are calculated with $K = 12$ classes, one class $C_i$ for each painter:

\begin{equation}
S_{w} = \sum_{i=1}^K S_i
\end{equation}

\begin{equation}
S_{b} = \sum_{i=1}^K N_i(\vec{\mu_i} - \vec{M})(\vec{\mu_i} - \vec{M})^T
\end{equation}

\noindent with $N_p$ the number of paintings in class $C_p$ and the
scatter matrix for class $C_i$ defined as

\begin{equation}
S_i = \sum_{i \in C_i} (\vec{f_i} - \vec{\mu_i})(\vec{f_i} - \vec{\mu_i})^T
\end{equation}

\noindent where $\vec{f_i}$ is an object of the feature matrix $F$ whose rows
and columns correspond to the paintings and its features $F = \left[ \leftarrow
  f_i^T \rightarrow \right]$ and $\vec{\mu_p}$ and $\vec{M}$ are the mean
feature vectors for objects in class $C_p$ and for all the paintings,
respectively:

\begin{equation} 
\vec{\mu_p} = \frac{1}{N_p} \sum_{i \in C_p} \vec{f_i}
\end{equation}

\begin{equation}
\vec{M} = \frac{1}{N} \sum_{i=1}^N \vec{f_i}
\end{equation}

The trace of within- and between-class ratio can be used to quantify dispersion:

\begin{equation} \label{eq:alpha}
\alpha = \mathrm{tr}(S_{b} S_{w}^{-1})
\end{equation}

Large values of $\alpha$ reveal larger dispersion and the features which relate
with large values of $\alpha$ are selected for the analysis
(Section~\ref{sec:two}).

\section{\label{sec:results} Results and discussion}

\subsection{Best features} \label{sec:two}

\begin{figure}[h!]
\begin{center}
        \includegraphics[scale=.5]{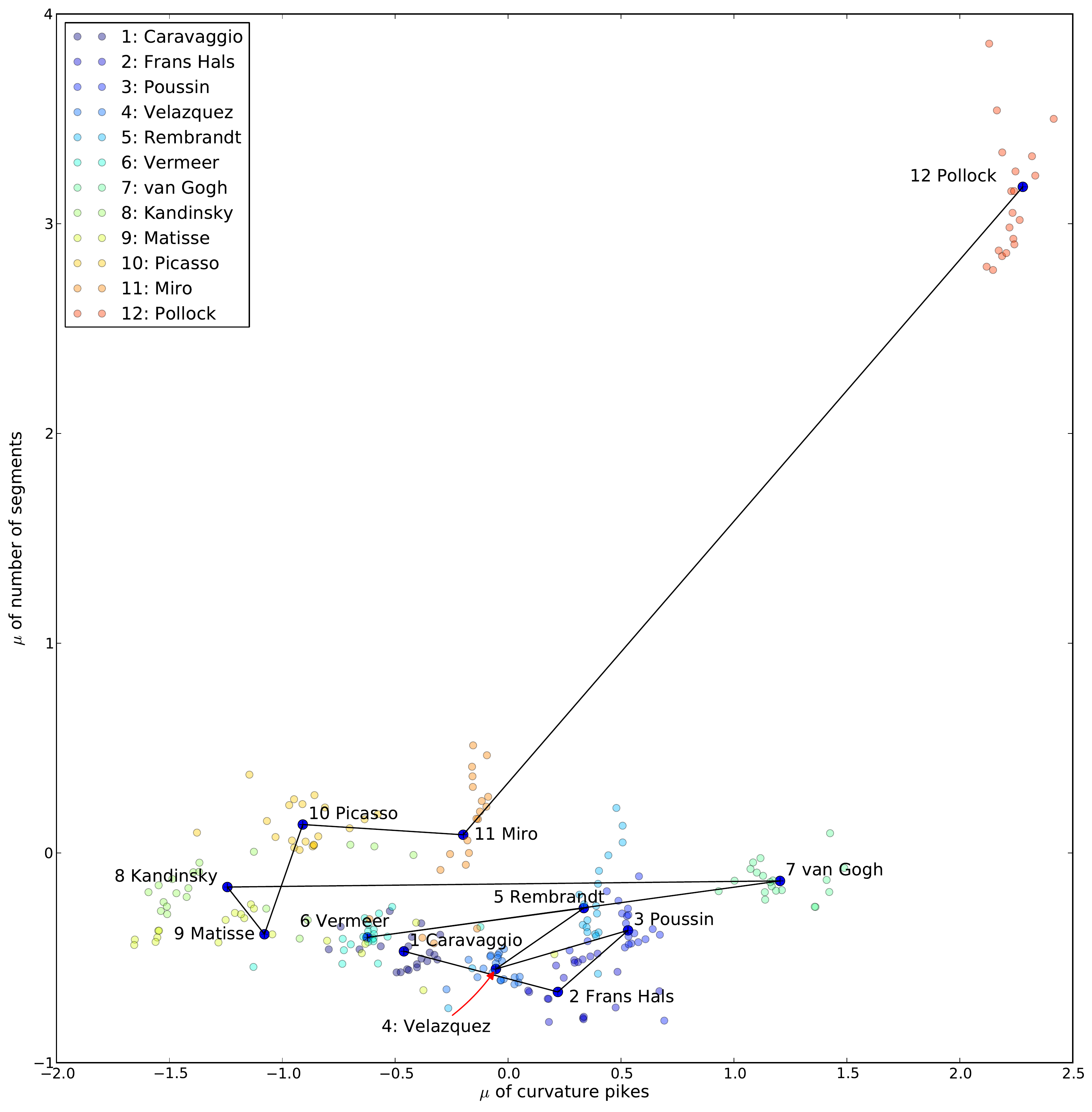}
      \caption{Projected \textit{painting space} considering the best pair of
        features: \emph{$\mu$ of curvature peaks} and \emph{$\mu$ of number of
          segments}.}
        \label{fig:caso1_g1}
        \end{center}
\end{figure}

By calculating $\alpha$ using Eq.~\ref{eq:alpha} for all possible feature
pairs $F_{N, a}$ and $F_{N, b}$ of the $N = 93$ features and ordering the
results by $\alpha$, it is possible to select the features which are most
relevant to classification: pairs with high $\alpha$ present better dispersion
and clustering than pairs with lower values. As shown in Table~\ref{tab:alpha}
(and Figure~\ref{fig:caso1_g1}), features \emph{$\mu$ of
  curvature peaks} and \emph{$\mu$ of number of segments} have the higher
$\alpha$ and are selected to opposition, skewness and dialectics analysis ---
both features are shown as predominant also in LDA, discussed in next
section. It is interesting to note the nature of selected features: the number
of segments and curvature peaks are the most prominent characteristics for the
classification of paintings, even better than texture or image complexity. Other
features presenting large values of $\alpha$ --- like $\mu$ of convex-hull area,
segments perimeter and area, and circularity --- are also related with shape
characteristics. Both features presented a similar projection and clustering
properties of Figure~\ref{fig:caso1_g1} as showed in
Figure~\ref{fig:scatters}.

\begin{table}[ht] \footnotesize
  \caption{\label{tab:alpha} Feature pairs $F_{N, a}$ and $F_{N, b}$ ordered by
    $\alpha$. Pairs with higher $\alpha$ present better dispersion and
    clustering. The best feature pairs \emph{$\mu$ of curvature peaks} and
    \emph{$\mu$ of number of segments} are selected for analysis and metrics
    calculation.}
\begin{indented}
\item[] \begin{tabular}{@{}llll}
\br

 Pair nr. & Feature $a$    & Feature $b$   & $\alpha$ \\ 
 
 \mr
 
 1 & $\mu$ of curvature peaks & $\mu$ of number of seg. & 42.445 \\
 2 & $\mu$ of number of seg. & $\mu$ of convex-hull area & 37.406 \\
 3 & $\mu$ of segments perimeter & $\mu$ of number of seg. & 36.703 \\
 4 & $\mu$ of segments area & $\mu$ of number of seg. & 36.214 \\
 5 & $\mu$ of number of segments & $\mu$ convex / original & 34.885 \\
 6 & $\mu$ of circularity ($\mathrm{Per.}^2/\mathrm{Area}$) & $\mu$ of number of seg. & 33.540 \\
 7 & Energy $\mu$ of image rows (green) & $\mu$ of number of seg. & 32.954 \\
 8 & Energy $\mu$ of rows and columns (green) & $\mu$ of number of seg. & 32.954 \\
 9 & Energy $\sigma$ of image rows (green) & $\mu$ of number of seg. & 32.932 \\
 10 & Energy $\sigma$ of rows and columns (green) & $\mu$ of number of seg. & 32.906 \\
 11 & $\mu$ of local entropy (5-size window) & $\mu$ of number of seg. & 32.898 \\
 12 & Entropy (Haralick adj. 4) & $\mu$ of number of seg. & 32.898 \\
 13 & Entropy (Haralick adj. 3) & $\mu$ of number of seg. & 32.883 \\
 14 & Entropy (Haralick adj. 1) & $\mu$ of number of seg. & 32.874 \\
 15 & Entropy (Haralick adj. 2) & $\mu$ of number of seg. & 32.869 \\
 16 & Energy $\mu$ of image rows (r.) & $\mu$ of number of seg. & 32.865 \\
 \end{tabular}
\end{indented}
\end{table}

The projected \textit{painting space} considering all the paintings that are
``represented'' by $\vec{p_i}$ is presented in Figure~\ref{fig:caso1_g1} which
reveals well clustered groups with minor superposition, mainly for modern
paintings. The time-series $S$ formed by prototypes $\vec{p_i}$ of each painter into
the projected space is shown as well.

A striking result is the
high distance which Pollock stays when compared with the other painters: it is
a consequence of the lag number of segments present in works of Pollock (the
y-axis being the projection of this feature: \textit{$\mu$ of segments
number}). Therefore, both the x-axis (\textit{$\mu$ of curvature peaks}) and y-axis
are relevant to separate the baroque and modern art movements. It is possible to
note a separation between baroque and modern painters where the baroque
paintings are arranged in an overlapping group while the modern painters are
more clustered and separated from each other while covering a widely region of
the \textit{painting space}. This is confirmed by the history of art with modern
painters being more individualists in their styles while baroque painters are used to share aesthetic characteristics in their paintings. The same
observation arises when following the time-series, the difference between the
movements is clear: while baroque artists tend to present a
recurring pattern, an abrupt displacement separates Van Gogh --- the first modern
painter in the \textit{painting space} --- from the previous, and breaks the
cyclic pattern. Van Gogh, although located near the baroque painters and in the opposite
extreme of modern painters, represents a transition to the modern period and after him the
following vector displacements will continue to evolve until reaching its apex
with Pollock.

\begin{figure}[ht!]
    
\begin{center}
{\centering
        \includegraphics[width=\columnwidth]{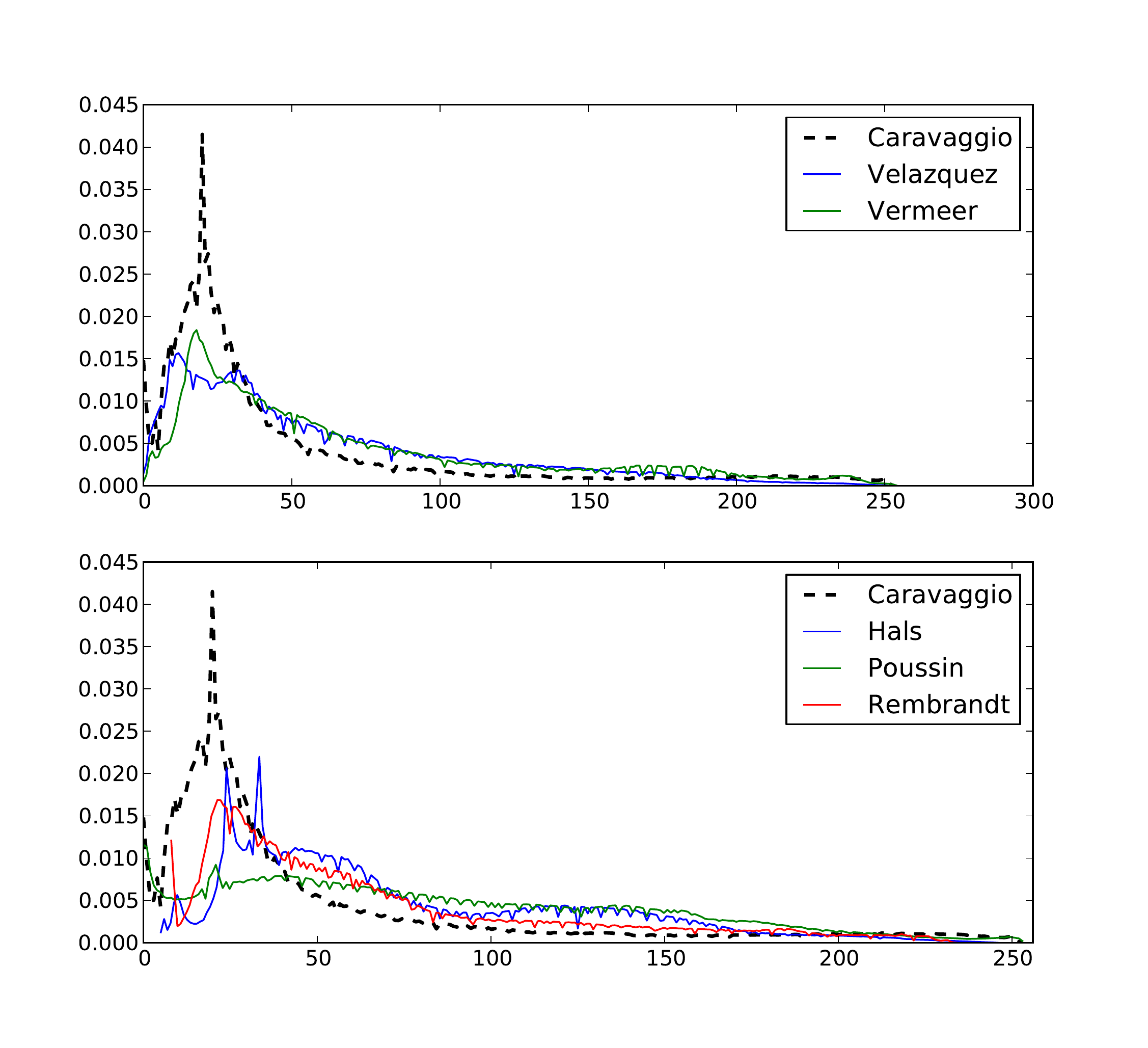}}
      \caption{Mean gray levels histograms for all the baroque painters. Vermeer and Velazquez show more similarity with Caravaggio than other baroque painters.}
        \label{fig:chiaroscuro}
  \end{center}
\end{figure}

While analyzing the baroque group separately, it is possible
to observe a trajectory drawn by Caravaggio and Frans Hals through
Poussin which ends with the opposite (and back forth)
movement of Vel\'{a}squez. It can be attributed to the influence of the
``\textit{chiaroscuro}'' master into these painters, mainly in Vel\'{a}squez who is known
to have studied the works of Caravaggio~\cite{gombrich}. It arises again in the
return to the Caravaggio movement by Vermeer -- some critics
affirm~\cite{lambert} that painters like Vermeer could not have even existed without
Caravaggio's influence: Vermeer and Caravaggio clusters are the most
superimposed considering all the portraits in the \textit{painting space}. Both facts are confirmed by the histograms of gray levels shown in Figure~\ref{fig:chiaroscuro}. Velazquez and Vermeer curves are more similar to Caravaggio than the remaining baroque painters.

\begin{figure}[ht!]
    
\begin{center}
{\centering
        \includegraphics[width=\columnwidth]{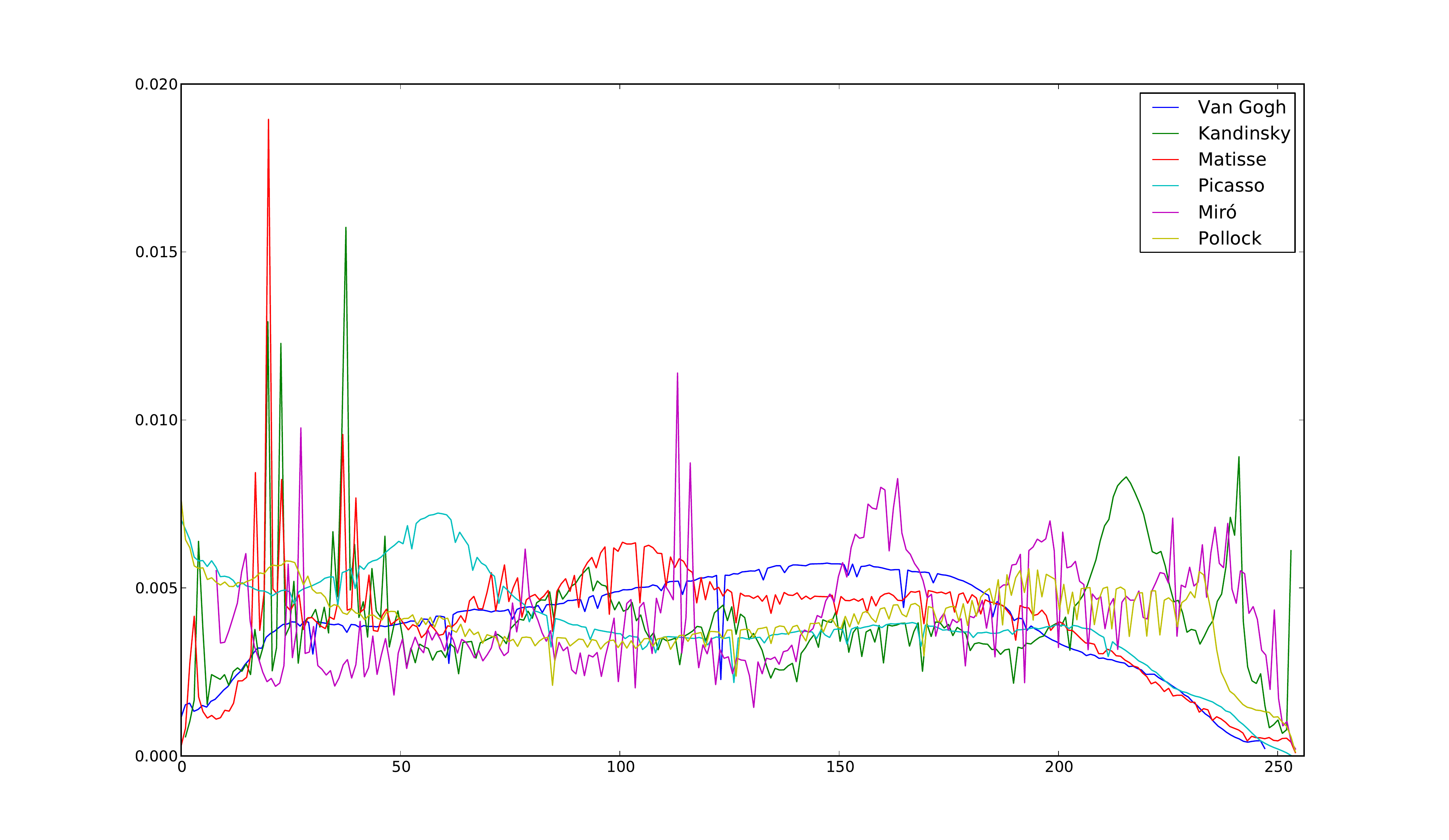}}
      \caption{Mean gray levels histogram for all the modern painters. There is minor similarities between modern artists.}
        \label{fig:chiaroscuro_modernos}
  \end{center}
\end{figure}

In summary,
the baroque group shows a strong inter-relationship by comparing with modern painters
where the absence of super-impositions is remarkable. Again, this suggests a strong
style-centric distinction among artists of the modern era while baroque artists
shared techniques and aesthetic characteristics. This is also confirmed when comparing the histograms of modern paintings in Figure~\ref{fig:chiaroscuro_modernos}: smaller similarities are observed between the considered artists, contrasting with baroque painters shown in Figure~\ref{fig:chiaroscuro}.

\begin{figure}[h!]
\begin{center}
        \includegraphics[width=\columnwidth]{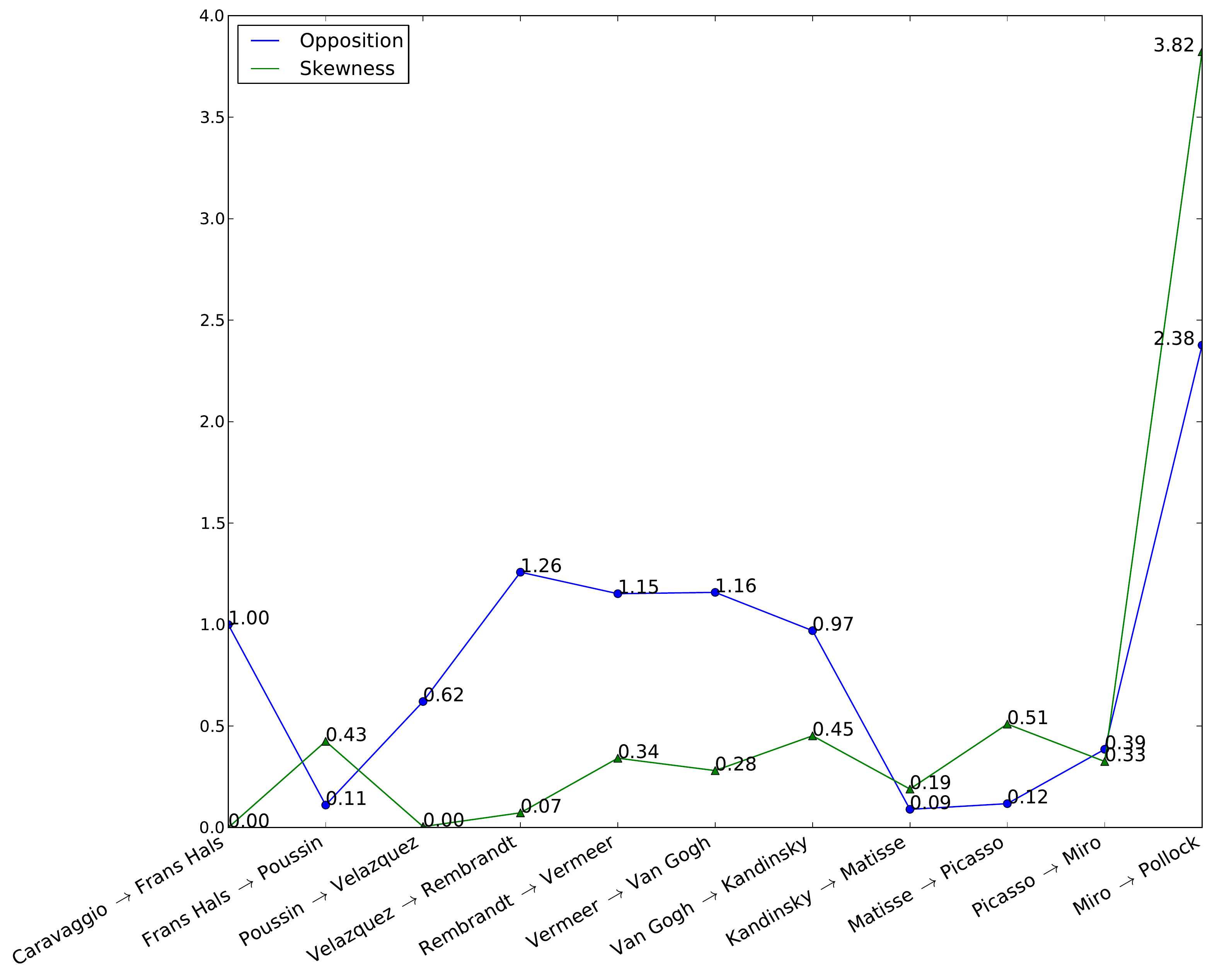}
      \caption{Opposition $W_{i,j}$ and skewness $s_{i,j}$ values for
        the two best features.}
        \label{fig:caso1_oposEinov}
        \end{center}
\end{figure}

When considering opposition and skewness, more interesting results
arise, as shown in Table~\ref{tab:opos1} and
Figure~\ref{fig:caso1_oposEinov}. Clearly, the larger value for opposition is attributed to Rembrandt. This is
surprising given that the Dutch master figures as a ``counterpoint'' of
baroque even being part of this art movement~\cite{gombrich}. Vermeer also
presents strong opposition and the nature of its paintings (e.g.\ domestic
interior, use of bright colors) could explain this phenomenon. A pattern is
shown in the beginning of baroque and modern art: an opposition decrease is
present in both cases, which is followed by an increase in
opposition. Henceforth, a following plateau of high opposition values is
observed in baroque painters. This plateau happens in the transition period
between baroque and modern art, gradually decreasing while the modern artists
begin to take place in history. This decreasing opposition values reflects a low
opposition role between first artists of baroque period and increasing
opposition as long the period is moving into modernism, although skewness values
remains oscillating and increasing during almost all the time-series. This
characterizes again a common scene in arts, mostly in modernists, each one
trying to define his own style and preparing to change into a new movement.  In
summary, the \textit{painting space} is marked by constantly increasing
skewness, strong opposition in specific moments of its evolution (the transition
between baroque and modern) and minor opposition between the artists of the same
movement.

\begin{table}[ht]
  \caption{\label{tab:opos1}Opposition and skewness indices for each
    of the twelve moves for a painter to the next.}
\begin{indented}
\item[] \begin{tabular}{@{}lll}
  
    \br
    Painting Move & $W_{i,j}$ & $s_{i,j}$ \\
    \mr
    Caravaggio $\to$ Frans Hals  &   1.     &  0.     \\
    Frans Hals $\to$ Poussin     &   0.111  &  0.425  \\
    Poussin $\to$ Vel\'{a}zquez   &   0.621  &  0.004  \\
    Vel\'{a}zquez $\to$ Rembrandt &   1.258  &  0.072  \\
    Rembrandt $\to$ Vermeer      &   1.152  &  0.341  \\
    Vermeer $\to$ Van Gogh       &   1.158  &  0.280  \\
    Van Gogh $\to$ Kandinsky     &   0.970  &  0.452  \\
    Kandinsky $\to$ Matisse      &   0.089  &  0.189  \\
    Matisse $\to$ Picasso        &   0.117  &  0.509  \\
    Picasso $\to$ Mir\'{o}       &   0.385  &  0.325  \\
    Mir\'{o} $\to$ Pollock       &   2.376  &  3.823  \\
    \br
  \end{tabular}
\end{indented}
\end{table}

\begin{figure}[h!]
\begin{center}
        \includegraphics[width=\columnwidth]{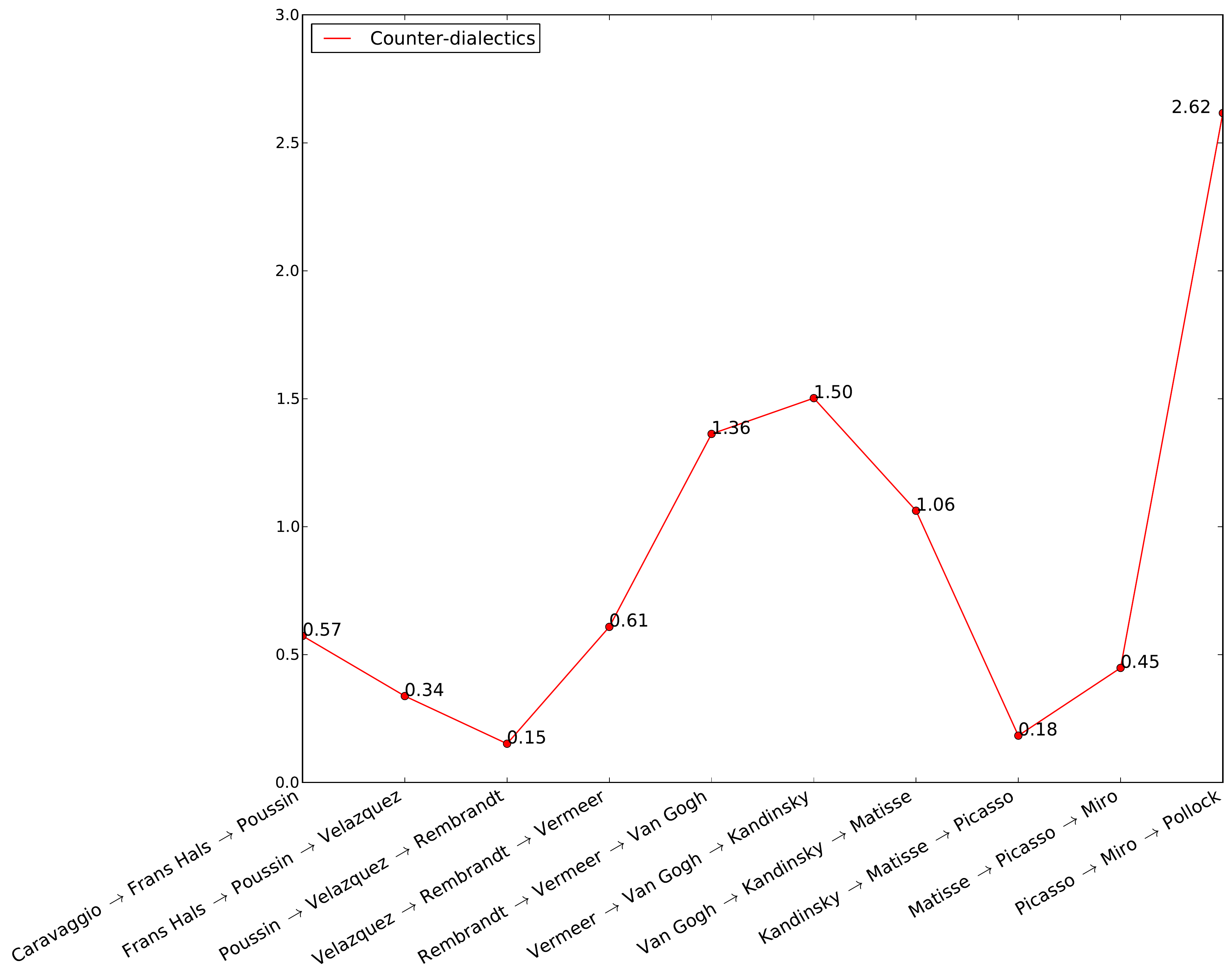}
    \caption{Counter-dialectics values considering the two best features.}
        \label{fig:caso1_dialetica}
        \end{center}
\end{figure}

The counter-dialectics, shown in Table~\ref{tab:dialetica1} and
Figure~\ref{fig:caso1_dialetica}, draws a parallel with the opposition and
skewness curves. It reinforces the already observed facts: painters of the same
movement show initially decreasing followed by increasing counter-dialectics
reflecting the concordance of members of the same movement and their preparation
to change into the next movement. The larger counter-dialectics happens in Van
Gogh and Kandinsky: again, the point where baroque ends and modern art starts,
regarding the painters selected for this study.

\begin{table}[ht]
  \caption{\label{tab:dialetica1} Counter-dialectics index for each of
    the ten subsequent moves among painters states for the best two features.}
\begin{indented}
\item[] \begin{tabular}{@{}ll}
  
    \br
    Painting Triple & $d_{i \rightarrow k}$ \\
    \mr
    Caravaggio $\to$ Frans Hals $\to$ Poussin   & 0.572 \\
    Frans Hals $\to$ Poussin $\to$ Vel\'{a}zquez & 0.337 \\
    Poussin $\to$ Vel\'{a}zquez $\to$ Rembrandt  & 0.151 \\
    Vel\'{a}zquez $\to$ Rembrandt $\to$ Vermeer  & 0.608 \\
    Rembrandt $\to$ Vermeer $\to$ Van Gogh      & 1.362 \\
    Vermeer $\to$ Van Gogh $\to$ Kandinsky      & 1.502 \\
    Van Gogh $\to$ Kandinsky $\to$ Matisse      & 1.062 \\
    Kandinsky $\to$ Matisse $\to$ Picasso       & 0.183 \\
    Matisse $\to$ Picasso $\to$ Mir\'{o}         & 0.447 \\
    Picasso $\to$ Mir\'{o} $\to$ Pollock         & 2.616 \\
    \br
  \end{tabular}
\end{indented}
\end{table}

\subsection{All the features}

\begin{figure}[h!]
\begin{center}
        \includegraphics[width=\columnwidth]{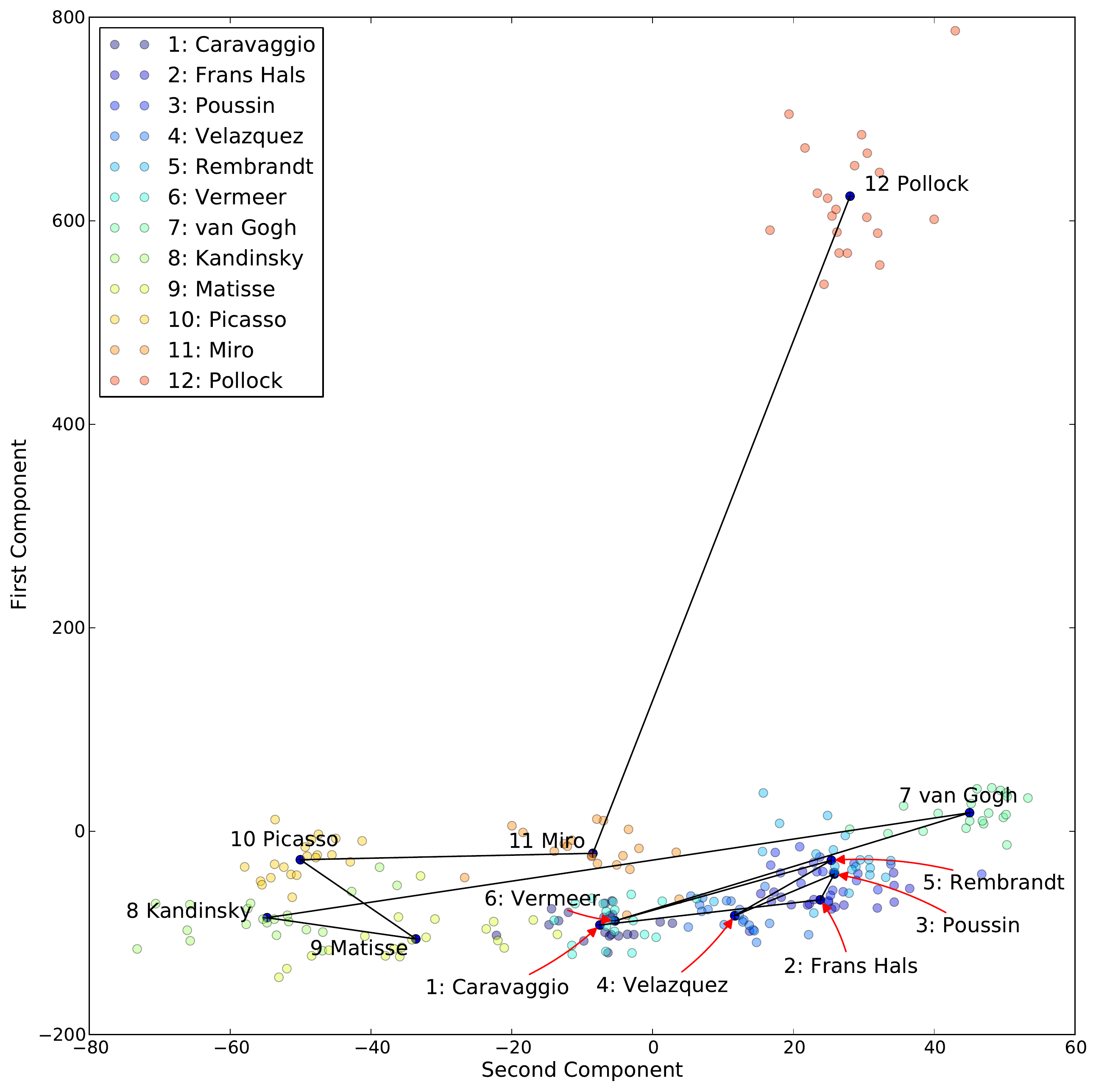}
      \caption{Time series yielded by 2-dimensional projected
        ``painting space'' considering the two first components
        obtained by LDA transformed into the $N = 93$ feature matrix.}
        \label{fig:caso3_g1}
        \end{center}
\end{figure}

\begin{figure}[h!]
\begin{center}
        \includegraphics[width=\columnwidth]{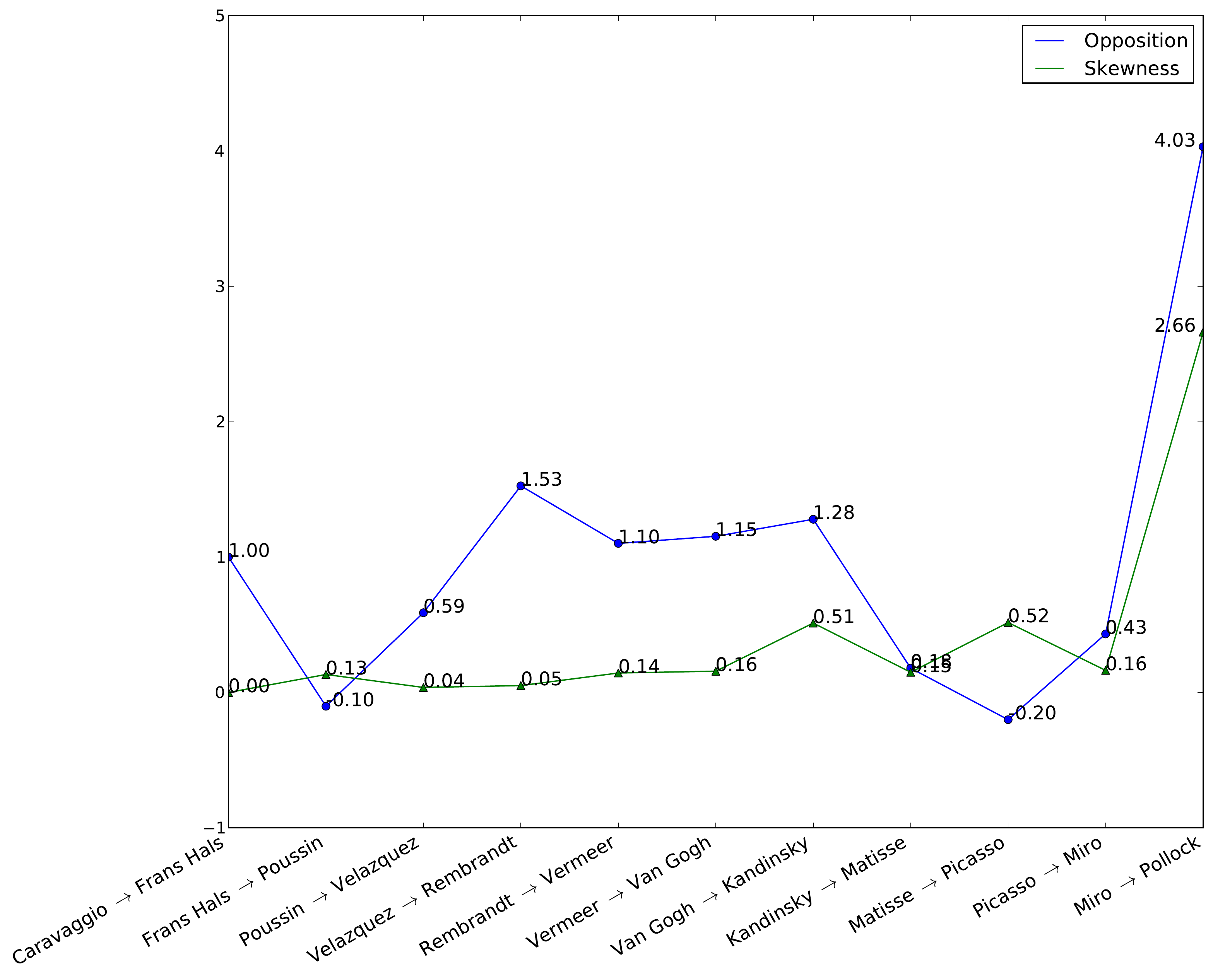}
      \caption{Opposition and Skewness values considering the time
        series for all the features. The same patterns observed when
        analyzing the best feature pair remains in this observation.}
        \label{fig:caso3_oposEinov}
        \end{center}
\end{figure}

\begin{figure}[h!]
\begin{center}
        \includegraphics[width=\columnwidth]{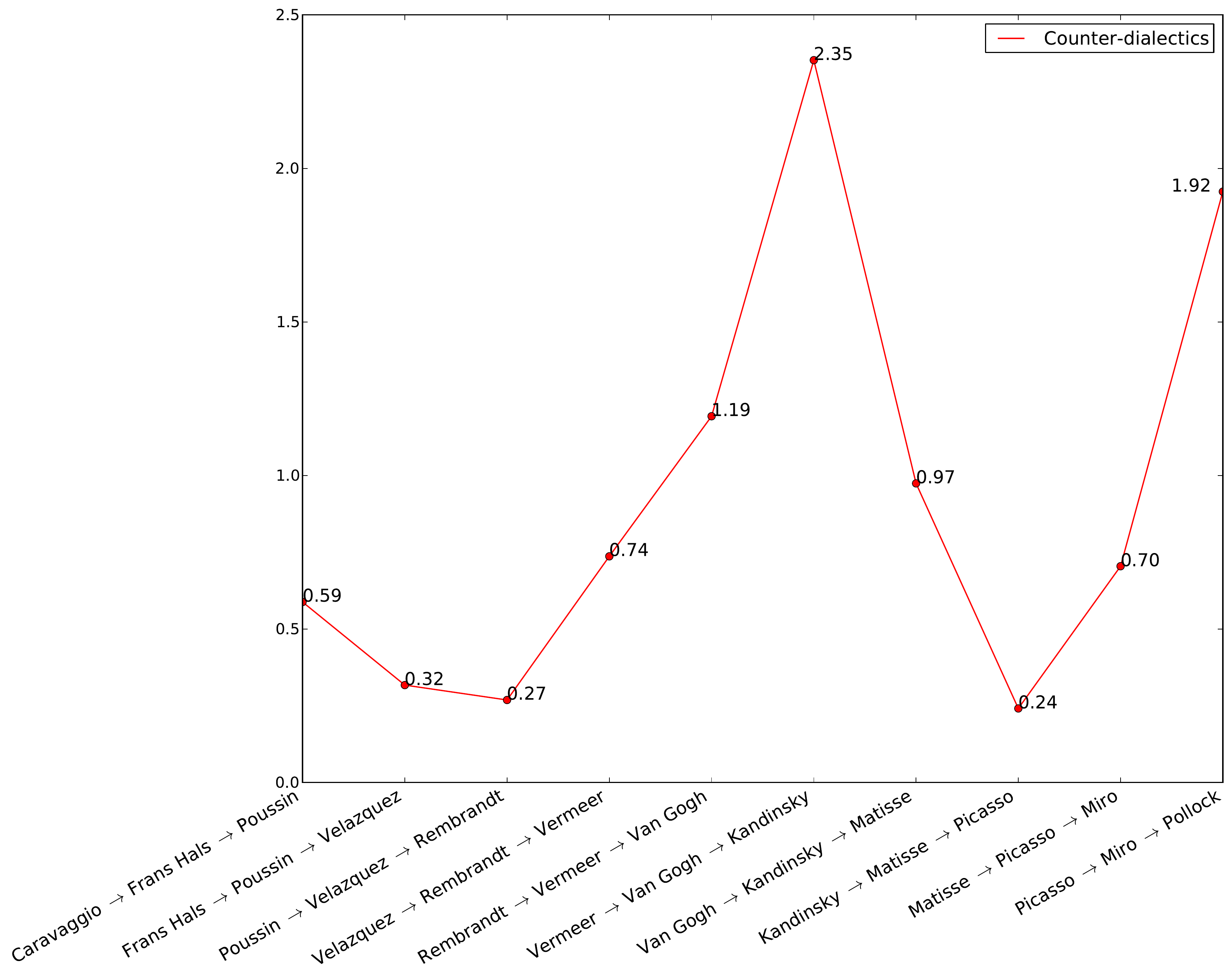}
      \caption{Counter-dialectics values (higher values reveals lower
        dialectics) considering all the features. The pattern observed
        in the best pair projection became stronger here: it is
        possible to observe clearly the highest value along the
        movement transition period (Van Gogh and Kandinsky).}
        \label{fig:caso3_dialetica}
        \end{center}
\end{figure}

Although features $F_{N, a}$ ($\mu$ of curvature peaks) and $F_{N, b}$ ($\mu$
of number of segments) showed as an interesting choice for classification, LDA
is applied considering all the $N = 93$ features to test the relevance of these
features and the stability of the results. The LDA method~\cite{luciano}
projected the features in a 2-dimensional space that better separates the
paintings and yields a time-series as done for the two most prominent
features. The first two components give the time-series shown in
Figure~\ref{fig:caso3_g1}. It is possible to note, as expected, a similarity
with results from Subsection~\ref{sec:two}. The skewness indices show even more
an ascending curve along the entire evolution, as presented in Table~\ref{tab:opos3} and Figure~\ref{fig:caso3_oposEinov}. The opposition and dialectics (Table~\ref{tab:dialetica3} and Figure~\ref{fig:caso3_dialetica})
patterns remain.

\begin{table}[ht]
\caption{\label{tab:opos3}Opposition and skewness indices for
    each of the twelve painters states moves}
\begin{indented}
\item[] \begin{tabular}{@{}lll}
    \br
    Painting Move & $W_{i,j}$ & $s_{i,j}$ \\
    \mr
    Caravaggio $\to$ Frans Hals  &   1.     &  0.     \\
    Frans Hals $\to$ Poussin     &  -0.101  &  0.132  \\
    Poussin $\to$ Vel\'{a}zquez   &   0.588  &  0.037  \\
    Vel\'{a}zquez $\to$ Rembrandt &   1.526  &  0.050  \\
    Rembrandt $\to$ Vermeer      &   1.101  &  0.143  \\
    Vermeer $\to$ Van Gogh       &   1.153  &  0.157  \\
    Van Gogh $\to$ Kandinsky     &   1.279  &  0.512  \\
    Kandinsky $\to$ Matisse      &   0.179  &  0.149  \\
    Matisse $\to$ Picasso        &  -0.201  &  0.516  \\
    Picasso $\to$ Mir\'{o}       &   0.432  &  0.163  \\
    Mir\'{o} $\to$ Pollock       &   4.031  &  2.662  \\
    \br
  \end{tabular}
  \end{indented}
\end{table}

\begin{table}[ht]
  \caption{\label{tab:dialetica3} Counter-dialectics index for each of
    the ten subsequent moves among painters states for the best two components
    of LDA projection.}
\begin{indented}
\item[] \begin{tabular}{@{}ll}
    \br
    Painting Triple & $d_{i \rightarrow k}$ \\
    \mr
    Caravaggio $\to$ Frans Hals $\to$ Poussin   & 0.587 \\
    Frans Hals $\to$ Poussin $\to$ Vel'{a}zquez & 0.317 \\
    Poussin $\to$ Vel'{a}zquez $\to$ Rembrandt  & 0.268 \\
    Vel'{a}zquez $\to$ Rembrandt $\to$ Vermeer  & 0.736 \\
    Rembrandt $\to$ Vermeer $\to$ Van Gogh      & 1.192 \\
    Vermeer $\to$ Van Gogh $\to$ Kandinsky      & 2.352 \\
    Van Gogh $\to$ Kandinsky $\to$ Matisse      & 0.974 \\
    Kandinsky $\to$ Matisse $\to$ Picasso       & 0.241 \\
    Matisse $\to$ Picasso $\to$ Mir'{o}         & 0.704 \\
    Picasso $\to$ Mir'{o} $\to$ Pollock         & 1.924 \\
    \br
  \end{tabular}
\end{indented}
\end{table}

\begin{figure}[h!]
\begin{center}
        \includegraphics[width=\columnwidth]{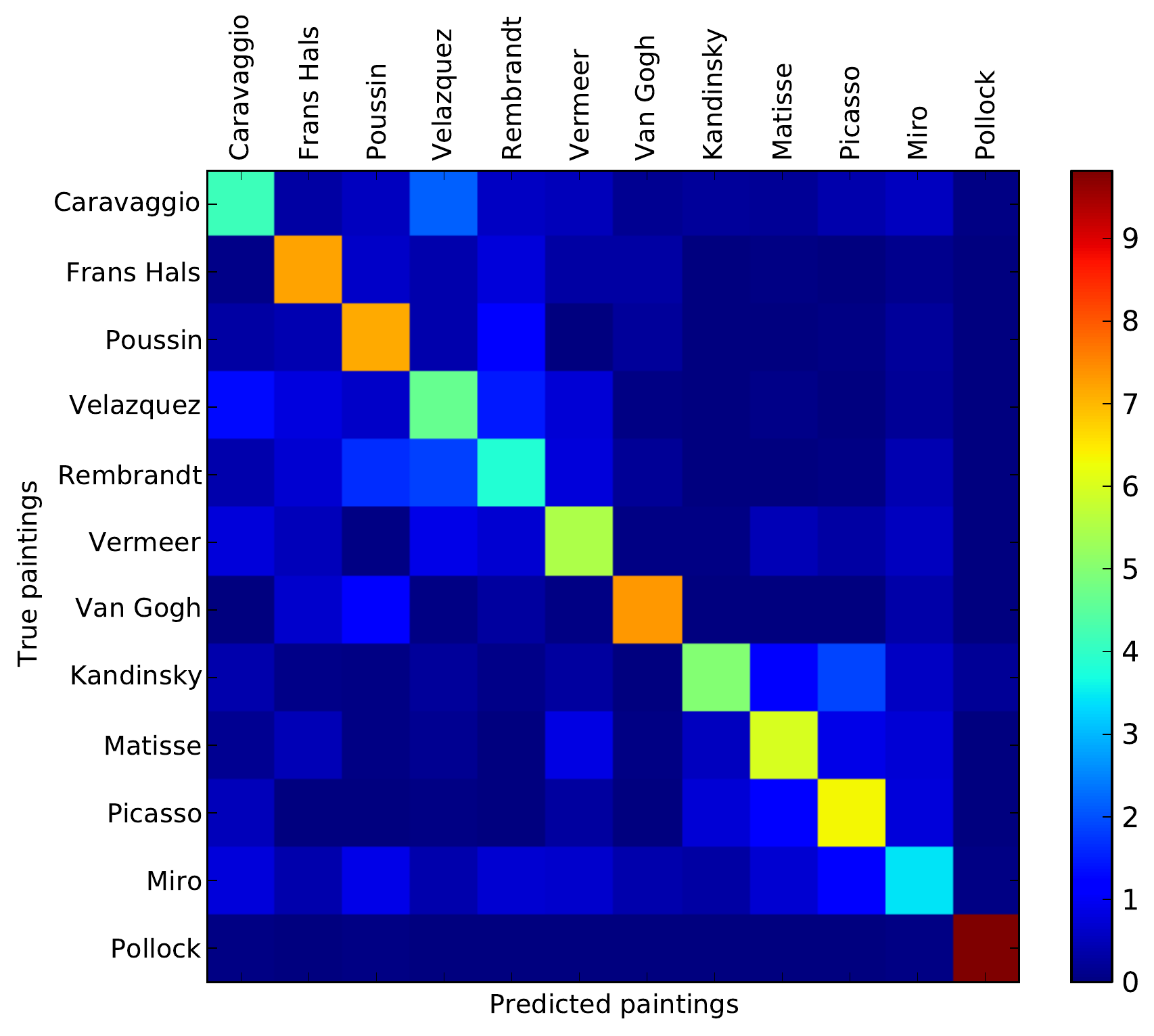}
      \caption{Confusion matrix for LDA. The half of paintings are used as a
        training set and the other half as test set. The validation is performed
        $100$ times. Diagonal elements shows the mean number of paintings in the
        predicted class (a painter) which equals to the true class.}
        \label{fig:cm}
        \end{center}
\end{figure}

For LDA validation, the total set of paintings is split in two groups: a
training set with 10 random selected paintings for each artist and a test set
with the remaining 10 paintings for each artist, without repetition. Such a validation is performed
$100$ times. The confusion matrix (Figure~\ref{fig:cm}) reveals the quality of
predicted output. Diagonal elements represent the mean number of samples for
which the predicted class is equal to the true class, while off-diagonal
elements indicates those ones that are unclassified by LDA. Higher diagonal
values indicate more correct predictions. As observed, the LDA method performed
as expected for the considered set of paintings. The best classified samples
are Pollock paintings which is expected given the high detachment of this
cluster observed in the presented projections. In general, the confusion matrix reflects facts previously discussed: a similarity between baroque painters, mainly Velazquez, Caravaggio and Rembrandt and a separation between painters before and after Van Gogh which defines the frontier between the baroque and modern movements.

\section{Conclusions}

It is shown that two features: \textit{a)} number of curvature peaks and
\textit{b)} number of segments of an image --- both related with shape
characteristics --- can be used for the classification of the selected painters
with remarkable results, even when compared with canonical feature measures like
Haralick or image complexity. Such relevance is supported by the analysis of a
dispersion index calculated for every pair of features and reinforced by LDA
analysis.

The effective characterization of selected paintings by means of these features allowed the definition of a ``painting space''. While represented as states in this projected space, the baroque paintings are shown as an overlapped cluster. The modern paintings clusters, in contrast, present minor overlapping and are disposed more widely in the projection. Those observations are compatible with the history of Art: baroque painters shared aesthetics while modern painters tended to define their own styles individually~\cite{gombrich}.  

A time-series --- composed by prototype states representing each painter chronologically --- allowed the concepts of opposition, skewness and dialectics to be approached quantitatively, as geometric measures. The painting states show a decrease in opposition and dialectics considering the
first members of the same movement (baroque or modern) followed by increasing
opposition and dialectics until it reaches the strong opposition momentum
between the two movements. Also, the skewness curve increases during
almost entire time-series. This could reflect a strong influence role of a
movement in its members together with an increasing desire to innovate, present
in each artist, stronger in modernists.

Both opposition, skewness and dialectics
measurements can be compared with results
already obtained for music and philosophy~\cite{vieira}. Music composers seems
to be guided by strong dialectics due to the recognized master-apprentice
role. Philosophers movements, otherwise, are strong in opposition. Painters, as this study reveals, show increasing skewness and strong opposition and counter-dialectics in specific moments of history.

While not sufficient to exhaust all the characteristics
regarding an artist or its work, this method suggests a framework to the study
of arts by means of a feature space and geometrical measures. As a future work, the number of painters could be increased and a set of
painters could be specifically chosen to analyze influence (e.g.\ works of Frans
Hals sons can be included to verify the influence of their father and master, or
paintings by Rafael, Poussin and Guido Reni~\cite{gombrich} or Carracci can be
compared to confront the already known similarity of both painters). A larger
number of paintings for each artist could be considered to analysis as well. The
same framework can be applied to other fields of interest like Movies or
Poetry. Another interesting use of this framework --- being currently developed
by the authors --- is a component of a generative art model: geometrical
measures in the \textit{painting space} (like the already defined dialectics or
opposition and skewness) can guide an evolutionary algorithm, assigning the
value of measures as the fitness of generated material. This model complements a
framework to the study of creative evolution in arts.

\appendix

\section*{Appendix}
\setcounter{section}{1}

\begin{figure}[h!]
\begin{center}
        \includegraphics[width=\columnwidth]{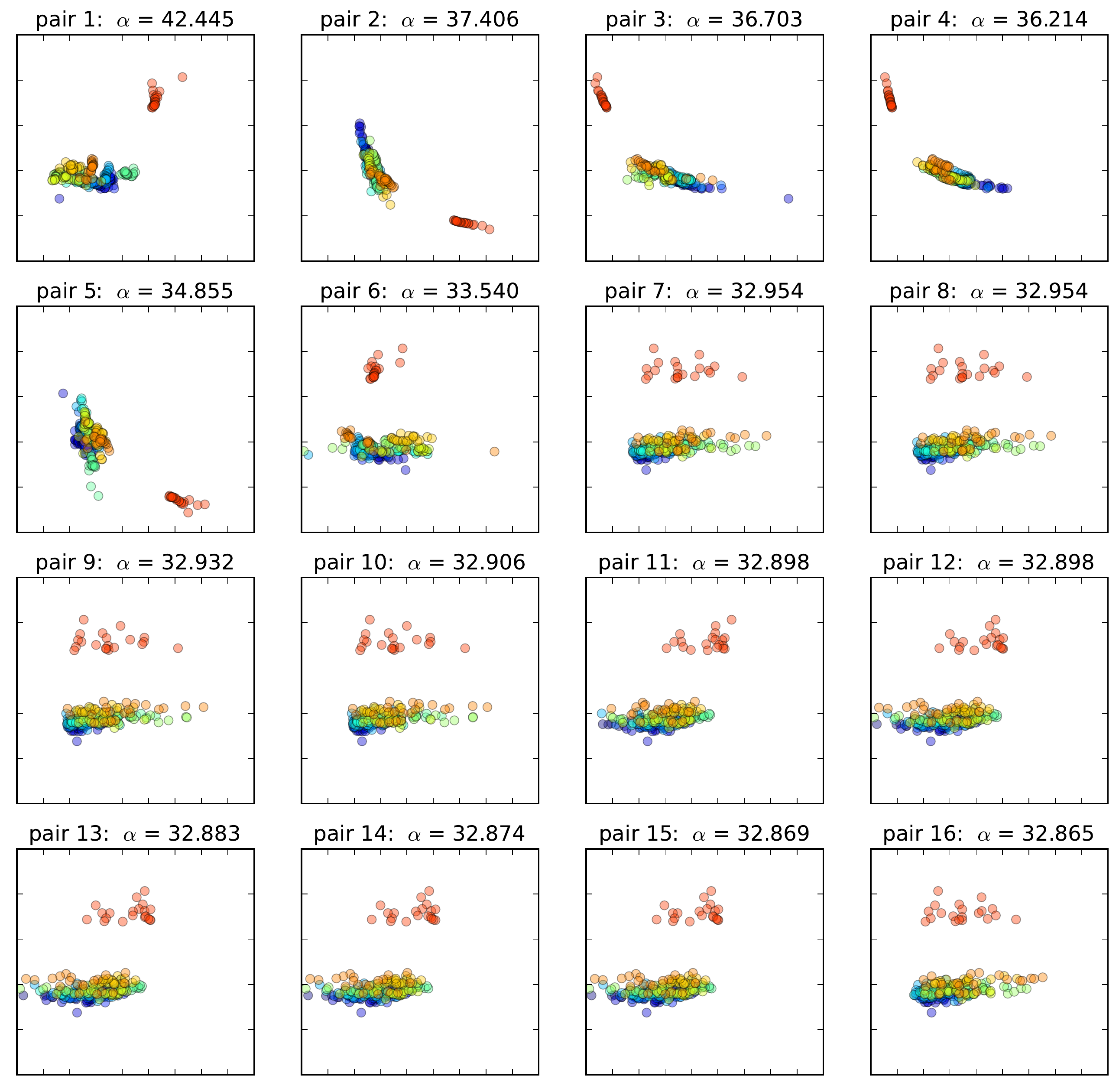}
      \caption{Scatter plots for each feature pair $i$ listed in
        Table~\ref{tab:alpha} with large values of $\alpha$. The first
        projection (pair $1$) was used for the analysis, however other
        projections (pairs $2 \ldots 16$) can be used.}
        \label{fig:scatters}
        \end{center}
\end{figure}

Although the first features pair ($\mu$ of curvature pikes and $\mu$ of number
of segments) is selected to the analysis, other features with large $\alpha$
values can be used as shown in Figure~\ref{fig:scatters}.

\bibliographystyle{unsrt}
\section*{References}
\bibliography{ana-pintores}

\end{document}